\documentclass[amsmath,amssymb,aps,floats,prl,twocolumn,showpacs,unsortedaddress,superscriptaddress]{revtex4}
\usepackage{graphicx}
\usepackage{amssymb}
\usepackage{epstopdf}
\usepackage{color}

\newcommand{\ie}{\emph{i.e.} }

\newcommand{\eff}{\mathrm{eff}}
\newcommand{\be}{\begin{eqnarray}}
\newcommand{\ee}{\end{eqnarray}}
\newcommand{\bfig}{\begin{figure}}
\newcommand{\efig}{\end{figure}}

\begin{document}
%\preprint{} 

\title{Large enhancement of emergent magnetic fields in MnSi with impurities and pressure}%

\author{Benjamin J.  Chapman}
\affiliation{Department of Physics, University of Colorado, Boulder, CO 80309, USA}%
\author{Maxwell G. Grossnickle$^\dagger$}
\affiliation{Department of Physics, University of Colorado, Boulder, CO 80309, USA}%
%\altaffiliation[Now at] {Department of Physics, University of California, Riverside, CA 92521}
\author{Thomas Wolf}
\affiliation{Institute for Solid-State Physics, Karlsruhe Institute of Technology, D-76021 Karlsruhe, Karlsruhe, Germany}%
\author{Minhyea Lee}
\email{minhyea.lee@colorado.edu}
\affiliation{Department of Physics, University of Colorado, Boulder, CO 80309, USA}%

\date{\today}

\begin{abstract}
We report a study of the topological Hall effect (THE) in  Fe-doped MnSi and compare with results from pure MnSi under pressure. We find that Fe doping increases the THE, indicating an enhancement of the magnitude of the emergent gauge field.  This is consistent with the concurrent reduction in the length scale of the skyrmion lattice. For both pressurized and doped samples, we calculate the emergent magnetic field based on the size of the measured THE, and compare it with a theoretical upper-bound.  We find that the ratio of these two remains more or less constant with pressure or Fe doping, but differs greatly from that of pure MnSi at ambient pressure. We discuss the implications of this ratio with respect to  trends in the saturated magnetic moment and helical pitch length as $T_C\rightarrow 0$ via doping and pressure, respectively.

%We present the  ratio between the magnitudes of the observed effective magnetic field and of the theoretical upper bound in both Fe doping and  both pressure that is   more or less constant across the pressure and Fe doping samples. We discuss the implication of this ratio with relation to distinct progress in the saturated magnetic moments and the helical pitch lengths as $T_C\rightarrow 0$ via doping and pressure, respectively.

%Transverse transport measurements provide a unique sensitivity to the existence of broken time-reversal symmetry, as manifested in e.g. the anomalous Hall effect of ferromagnets, or the edge state of quantum Hall systems.  This sensitivity can give direct access to gauge fields of purely quantum-mechanical origin.  Here, we observed a largely enhanced topological Hall effect (THE) in Mn$_{1-x}$Fe$_x$Si, arising from a gauge field generated by the material's novel spin texture.  Experimental estimates of this emergent gauge field are made with the measured THE for both pure MnSi under pressure and Fe-doped MnSi at ambient pressure, and compared to upper-bounds from theoretical considerations.

\end{abstract}

\maketitle
A new type of magnetically-ordered state was recently discovered  \cite{Muhlbauer:2009,Munzer:2010,Yu:2010,Seki:2012,YuRS:2010} to form a lattice from vortex-like objects in the spin texture, so-called skyrmions \cite{Bogdanov:1989}. 
Two-dimensional skyrmion lattices (SLs) are stabilized in the presence of an external magnetic field and thermal fluctuations near the Curie temperature ($T_C$), occupying a small pocket  in the temperature ($T$) - applied magnetic field ($B$) phase diagram (the $A$-phase). They orient perpendicular to the applied field $B\hat{z}$, and have been observed with small angle neutron scattering (SANS) \cite{Muhlbauer:2009,Munzer:2010} and Lorentz tunneling electron microscopy \cite{Yu:2010,Seki:2012,YuRS:2010}, exclusively in B20-structures with cubic symmetry (space group $P2_13$). Celebrated examples  include  the itinerant ferromagnet MnSi \cite{Muhlbauer:2009}, semiconductors  Fe$_{x}$Co$_{1-x}$Si \cite{Munzer:2010,YuRS:2010} and FeGe \cite{Yu:2010}, and insulator Cu$_2$SeO$_3$\cite{Seki:2012, Adams:2012}. The B20 crystal structure lacks inversion symmetry, allowing a new energy scale ($D$), the Dzyaloshinsky-Moriya interaction \cite{Moriya:1960,Dzyaloshinsky:1958}, to compete with the conventional exchange interaction ($J$). This results  in a helically ordered magnetic ground state with well defined pitch $\lambda$, whose length is proportional to $J/D$.  The length scale for the unit cell of an SL is also set by  $\lambda$ \cite{PfleidererJPCM:2010}, which runs from  a few nm to a few tens of nm, depending on the compound. 

The non-coplanarity of spin structures like hexagonal SLs can give rise to an emergent magnetic field $\boldsymbol b_r$ expressed \cite{Bruno:2004,Schulz:2012} as
\be
b_r^i = \frac{\Phi_0}{8 \pi} \epsilon_{ijk}\hat{n} \cdot(\partial_j \hat{n} \times \partial_k \hat{n}).
\label{skdens}
\ee
Here $\epsilon_{ijk}$ is the Levi-Civita symbol whose indices run over $x,y$, and $z$, $\hat{n}(\boldsymbol {r})$ is a unit vector of the magnetization $\boldsymbol {M}(\boldsymbol {r})$, and $\Phi_0 = h/|e|$ is the single-electron flux quantum.  The hexagonal SL in MnSi is uniform in the $\hat{z}$ direction, so only the $z$-component of $\boldsymbol{b_r}$ is non-vanishing: $\boldsymbol{b_r} = (0,0,b_r)$.  This field arises in the strong Hund coupling limit, where the spin of conduction electrons orients parallel to the local magnetization, twisting to follow it as they move through the material. This results in the acquisition of an extra phase factor in their wave functions, represented by the line integral of a vector potential, analogous to the Aharonov-Bohm effect \cite{Aharanov:1959}.  The curl of that vector potential is the quantity given in Eq.~\ref{skdens}, which acts on conduction electrons in a similar way as the physical magnetic field \cite{Ye:1999, Ioffe:1996}, and causes the topological Hall effect (THE).% in addition to the contributions from the normal and anomalous contributions generated by  the applied field and the spontaneous magnetic moment, respectively.

Similarly,  the intrinsic  anomalous Hall effect in ferromagnets is caused by a fictitious k-space magnetic field $\boldsymbol b_k$  (see~\cite{Nagaosa:2010} and references therein) that arises from a Berry-curvature in momentum space.  In fact, it was recently suggested \cite{Ritz:2013} that chiral magnets playing host to SLs may be the first experimental examples of systems in which mixed-Berry phases, that is, Berry phases acquired from closed orbits in the full 6-dimensional phase-space ($\boldsymbol r, \boldsymbol p$), may play an important role.
 
%There have been  reports on peculiar Hall effects observed in more complicated spin structures of  frustrated  pychlore \cite{Taguchi:2001} or frustrated triangular lattices \cite{Takatsu:2010}, in which a complex magnetic field dependence was attributed to a real space spin texture, \ie~a magnetization sufficiently varied to generate a non-zero topological field $\boldsymbol{b_r}$. It is difficult, however, to distiguish this from other contributions to the Hall effect, and hence challenging to identify the existence of such a field in these compounds. On the other hand, the THE observed in pure MnSi \cite{Neubauer:2009} is characterized by a unique $T$ and   $B$ dependence, and a potentially enormous magnitude, as with MnSi under hydrostatic pressure \cite{Lee:2009}. Moreover, the emergence of such a signal is well confined in the $T-B$ plane to a region known as the $A$-phase~\cite{Thessieu:1997}, where a skyrmion lattice was indeed detected in small angle neutron scattering (SANS) experiments \cite{Muhlbauer:2009}.   

In an SL, the area integral of $\boldsymbol{b_r}$ over a magnetic unit cell is quantized to an integer times $\Phi_0$ \cite{Bruno:2004,Binz:2008}.  For MnSi that integer is $-1$ \cite{Muhlbauer:2009}, and the similarity in SANS data of 8\% Fe doped samples  \cite{PfleidererJPCM:2010} suggests this topological quantum number is unchanged with the level of doping considered here: %removed 'to pure MnSi' 
\be
\oint_{\mathcal A_{sk}}  \boldsymbol{b_r} \cdot \; \mathrm d\boldsymbol{\mathcal A} =  -\Phi_0,   
\label{Beff}
\ee
where $\mathcal {A}_{sk} = (2/\sqrt{3}) \lambda^2$ is the area of the magnetic unit cell and the minus sign implies that the emergent field on average opposes the normal vector of the unit cell, which is parallel to the applied $B$.  Eq.~\ref{Beff} allows estimation of an \emph{upper bound} for  the average topological magnetic field induced by the spin texture: $\bar{b_r} =  - \Phi_0 / \mathcal A_{sk}$.  The size of this field is tunable via control of $\lambda$ and typically tens of Tesla in Mn$_{1-x}$Fe$_{x}$Si (See Table~\ref{t2}).
%Due to the small size of  $\mathcal A_{sk}$, $\bar{b_r}$ can reach a few tens of Tesla.% of effective field.%which may be utilized for application purposes \cite{skyrmionics}.

%\begin{figure}[htb]
%\begin{center}
%\includegraphics[width=0.6\linewidth]{BtopV2.eps}
%\caption {\small Spatial dependence of  $b_r^z(\boldsymbol r)$ for triple axis twisted helical spin texture \cite{Binz:2008}, modeled for pure MnSi. $+\hat z$ is parallel to $B$ and to the plane of skyrmion. The red arrows show the wave-vectors for the three helices that compose the triple-q structure in the magnetization.  The topological field is strongest (dark-blue stripes) where adjacent skyrmions touch.}
%\label{Btop}
%\end{center}
%\efig 

%This is in general a non-trivial calculation, but many of its details are contained in the normal Hall coefficient \cite{Neubauer:2009}.  A reasonable approximation for the effective gauge field felt by conduction electrons is then given by $ \rho_{yx}^T = R_H B_{\eff}$, where $R_H$ is the normal Hall coefficient and $B_{\eff}$ is some positive fraction of . 

\begin{table*}[htb]
\begin{tabular}{l || r || r |r | r|r| r| r }
\hline
	&      0\%  & 8 kbar & 6\% Fe& 10 kbar & 9\% Fe& 13 kbar& 13\% Fe\\
\hline
$T_C$ [K]& 29.5 $\pm$ 0.2 & 17.8 $\pm$ 0.3  & 14.5 $\pm$ 0.6  & 13.4 $\pm$ 0.2 & 8.5 $\pm$ 0.25  & 7.3 $\pm$ 0.1 & 5.5 $\pm$ 0.3\\
$m_{s}$ [$\mu_B$]~\cite{Koyama:2000} &0.43 $\pm$ 0.05&0.36 $\pm$ 0.2 &0.31 $\pm$ 0.05&0.3 $\pm$ 0.1 &0.22 $\pm$ 0.05&0.31 $\pm$ 0.1 &0.19 $\pm$ 0.05\\
\hline\hline
$R_H$ [n$\Omega$cm/T]     & +7.3 $\pm$ 0.1 \cite{Lee:2007} & +9.6 $\pm$ 1.6 & & +7.4 $\pm$ 1.5 & -2.0 $\pm$ 0.7  & +5.0  $\pm$ 0.5 & -1.0 $\pm$ 0.7  \\
$\rho_{yx}^T$ max [n$\Omega$cm]      & -4.5 $\pm$ 1 \cite{Neubauer:2009} & -38 $\pm$ 2 & +7 $\pm$ 1  & -43.7 $\pm$ 1.7  &+20 $\pm$ 1   & -31.4 $\pm$ 2.2 &  32 $\pm$ 2 \\
$\lambda$ [nm]~\cite{Grigoriev:2009,Fak:2005}     & 17.4 $\pm$ 0.4 & 15.2 $\pm$ 0.2 &12.4 $\pm$ 0.4  & 15.1 $\pm$ 0.2  & 10.4 $\pm$ 0.4  &  14.4 $\pm$ 0.1 & 7.4 $\pm$ 0.5 \\
\hline
$B_{\eff} = \rho_{yx}^T$/$R_H$ [T]&  -0.61 $\pm$ 0.14 & -3.9 $\pm$ 0.7 & &  -5.9 $\pm$ 1.3 & -10.0 $\pm$ 3.5    & -6.3 $\pm$ 0.8 & -30 $\pm$ 21  \\
$\bar{b_r}  = \Phi_0/\mathcal {A}_{sk}$ [T] & -11.9 $\pm$ 0.4 \footnotemark & -15.5 $\pm$ 0.2 & -23.1 $\pm$ 0.9  &  -15.7 $\pm$ 0.2  & -33.3 $\pm$ 1.7  &  -17.1 $\pm$ 0.1 & -66 $\pm$ 6    \\
$f = B_{\eff}/\bar{b_r}$	        & 0.05 $\pm$ 0.01  & 0.25 $\pm$ 0.04 &   &  0.38 $\pm$ 0.08 &  0.30 $\pm$ 0.11 &   0.36 $\pm$ 0.04 & 0.45 $\pm$ 0.32 \\
\hline
\end{tabular}
 \caption {\small   The effects of Fe doping and pressure on the emergent magnetic fields in MnSi. Values for $\rho_{yx}^T$  are the maximum topological signal from $B$ sweeps.  $R_H$s were computed with fits to Eq.~\ref{AHEeq} for 0\%, 9\%, and 13\% Fe.  For MnSi under $P$ they were extracted by computing the slope of $\rho_{yx}$ when $B > B_p$.\\
%\vspace{3mm}
 \\
\footnotesize $^a$  This was estimated as 13.15 T in other work~\cite{NeubauerErratum:2013}, the discrepany arises from use of a slightly different $\lambda$ in $b_r$'s calculation. } %Magnetization values from Ref.~\cite{Koyama:2000} were taken at 4.2 K and 0.7 T.}
\label{t2}
\end{table*}

Hall effects arising from non-trivial spin textures \cite{Lee:2009,Neubauer:2009} provide a rare opportunity to directly access gauge fields of purely quantum-mechanical origin.  To connect  $\bar{b_r}$ to our experimental results, we introduce a ratio $f = B_{\eff}/\bar{b_r}$ between the theoretial upper bound $b_r$ and an effective field $B_{\eff}$ estimated from the magnitude of THE and the normal Hall coefficient $R_H$, \emph{viz.} $ \rho_{yx}^T = R_H B_{\eff}$ \cite{Ye:1999}.  

Motivated to tune the skyrmion size set by $\lambda$, and thus control the magnitude of $\bar{b_r}$ % via Eq.~\ref{Beff}, 
we studied the Hall effect in Fe doped MnSi. The primary effect of increasing Fe is a suppression of $T_C$ and the saturated magnetic moment ($m_{s}$) \cite{Grigoriev:2011,Bauer:2010}, both of which monotonically go to zero, vanishing at the critical doping $x_c \simeq 15-19\%$ \cite {Grigoriev:2009,Bauer:2010}.  Previous studies have shown that Fe doping leads to a linear decrease in $J$ but leaves $D$ unchanged \cite{Grigoriev:2009}, resulting in a decrease in $\lambda$ observed in scattering experiments \cite{Grigoriev:2009,PfleidererJPCM:2010}.  This compresses the spatial extent of the skyrmions, squeezing their single flux quantum through a smaller area, and necessitating a larger $\bar{b_r}$. 

To provide insight on how the suppression of $T_C$ affects emergent magnetic fields, we compare the above results with Hall measurements of pure MnSi under pressure \cite{Lee:2009,Ritz:2013}.  Like Fe doping, pressure ($P$) also suppresses $T_C$, reaching zero at the critical pressure $P_C \simeq 15$ kbar. $m_S$, however, is not as severely suppressed as in doped samples with comparable $T_C$s,  and even remains non-zero at $P_C$ \cite{Koyama:2000}. 
Furthermore, $\lambda$ is reduced only slightly under pressure: at $P_C$, $\lambda$ is $\approx$ 80\% of its length at ambient pressure ($\lambda_0$) \cite{Fak:2005}; it falls to 40\% of $\lambda_0$ with 13\% Fe doping.  We return to these differences in later comparisons of the THE.

We find that the topological Hall effect in Fe doped samples is enhanced by  factors of 2, 5, and 8 with respect to MnSi. The sign of the THE flips as well, remaining opposite to the sign of $R_H$, in accordance with the minus sign in Eq.~\ref{Beff}. %{\color{green} Pressure data (see also~\cite{Lee:2009,Ritz:2013}) is included for the purpose of a quantitative comparison to the effect of Fe doping.}  
$f = B_{\eff}/\bar{b_r}$ ranges between 0.25 and 0.45 for doped and pressurized samples, while for MnSi at ambient $P$ the ratio is much lower, only 0.05. We discuss later the implications of these disparate $f$ values.  Table~\ref{t2} summarizes our experimental results and calculations of the emergent fields in MnSi subjected to Fe doping and $P$.

Single crystals of MnSi and Mn$_x$Fe$_{1-x}$Si were grown by the Bridgman technique, and cut with typical dimensions $\sim$ 2 mm $\times$ 0.7 mm $\times$ 100 $\mu$m.  Six contacts of gold wire and silver paint were used for Hall and magnetoresistance measurements, for which DC currents of 0.5 -- 4mA were applied. All data presented here are for experiments with $B$ along the $\langle111\rangle$ direction, with the current perpendicular to $B$, and demagnetization effects corrected according to Ref.~\cite{Aharoni:1998}.  Typical residual resistivity ratios, determined by $\rho$(300K)/$\rho$(2K), were 80, 8, 5, and 4 for the 0\%, 6\%, 9\%, and 13\% Fe contents. Methods for the pressure experiments are described in~\cite{Lee:2009}.

Fig.~\ref{rhoM}(c) shows the $T$ dependence of the longitudinal resistivity $\rho$ at zero field. $T_C$s of 29.5, 14.5, 8.5, and 5.5 K were determined from the discontinuity in  d$\rho$/d$T$, consistent with  $M$ vs $T$ (data not shown) for 0\%, 6\%, 9\%, and 13\% respectively.  In general, higher Fe contents are more resistive.  Reduction of $m_S$ upon Fe doping is also obvious in Fig.~\ref{rhoM}(b), which shows $M$ as a function of $B$ taken at different fixed $T$s.  Panel (a) of Fig.~\ref{rhoM} shows  the resistivities normalized by  $\rho(T=T_C)$, as a function of $T/T_C$.  Note the obvious trend: as $T_C$ is suppressed by pressure or doping, the fractional change in resistivity is less sensitive to fractional changes in temperature.

\begin{figure}[htb]
\begin{center}
\includegraphics[width=0.6\linewidth]{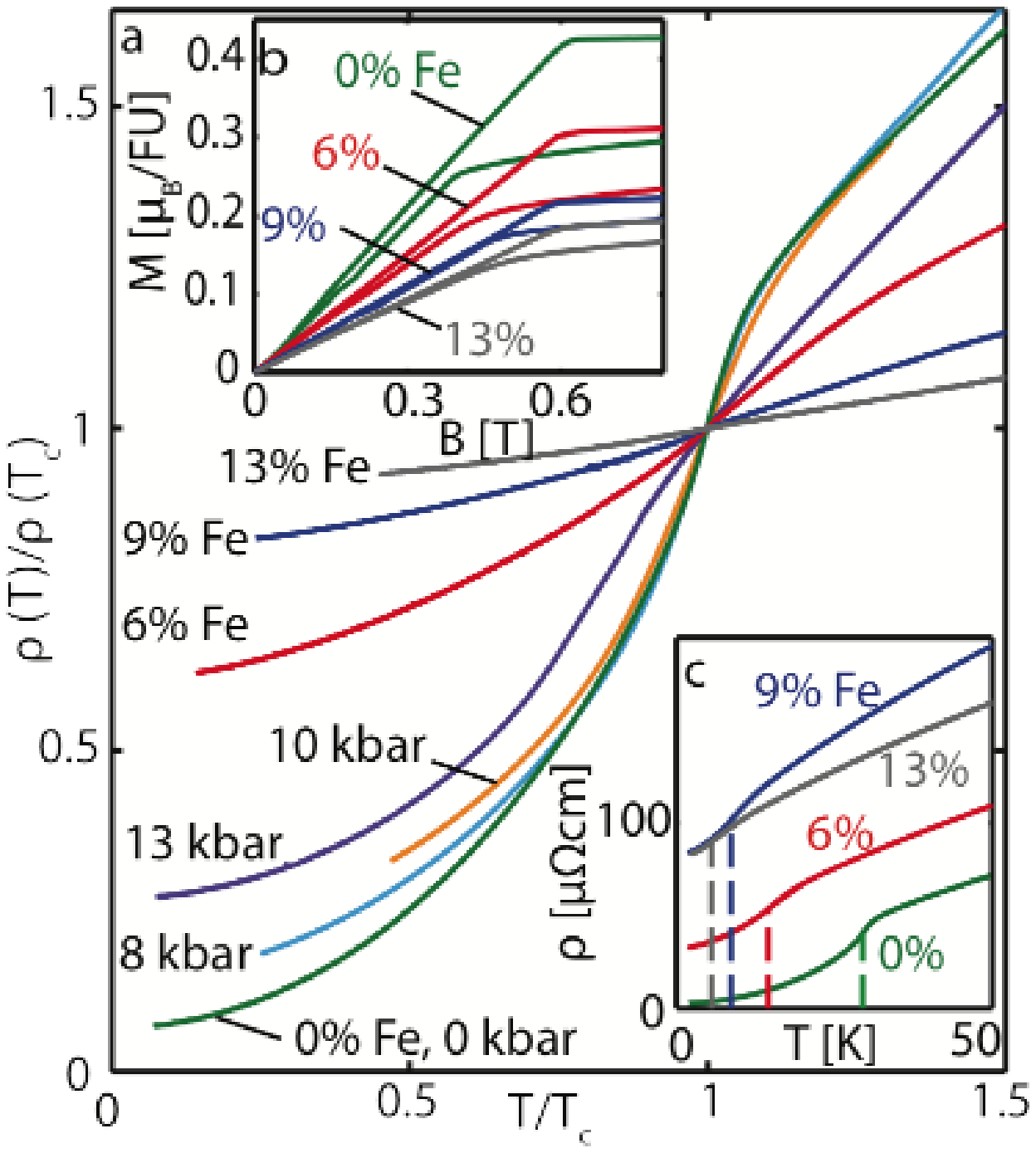}
\caption {\small 
(a): Normalized  resistivities $\rho(T)/\rho(T=T_C)$ as a function of $T/T_C$ for both Mn$_{1-x}$Fe$_x$Si and MnSi under $P$.  (b): $M$-$B$ curves for the doped samples %at fixed $T$s: 0.07, 0.41, 0.81, and 0.95 $T_C$ (0\%); 0.14, 0.41, 0.69, and 0.96 $T_C$ (6\%); 0.24, 0.53, and 0.82 $T_C$ (9\%); 0.38 and 1.1 $T_C$ (13\%).  
2K and $T_C$.  (c): $\rho(T)$ for 0\%, 6\%, 9\%, and 13\%, for which $T_C$s are 29.5 K, 14.5 K, 8.5 K, and 5.5 K (vertical lines). For MnSi under $P$, $\rho(T)$ is of comparable magnitude to the 0\% Fe curve in (b).}
\label{rhoM}
\end{center}
\efig

The transverse resistivity $\rho_{yx}$ is plotted as a function of $B$ in Fig.~\ref{Hall}. (a-c) show 6\%, 9\%, and 13\% Fe content.  For purposes of comparison, (d-f) show pure MnSi at 8, 10, and 13 kbar, pressures chosen to have comparable $T_C$s to the doped samples.  Note the $y$-axes' scales are different for each panel, with the bar indicating 20 n$\Omega$cm.  The unique field profile of $\rho_{yx}$ near $T_C$ clearly demonstrates a topological contribution $\rho_{yx}^T$, visible as a bulge within the narrow range $0.2<B<0.35$ T in both 6\% and 9\%, and broadening in the 13\% data. The maximum values of $\rho_{yx}^{T}$ for 6\%, 9\% and 13\% are about 2,  5 and 8 times bigger than the signal reported in MnSi at ambient $P$ \cite{Neubauer:2009}, consistent with an increasing $\bar{b_r}$ caused by a reduction of the magnetic unit cell's size.  The sign of $\rho_{yx}^T$ is positive for all doped samples.  This can be contrasted with MnSi under $P$~(see Table~\ref{t2}), where THE is always negative and of even larger magnitude, consistent with previous reports~\cite{Lee:2009,Ritz:2013}. 

While pure MnSi displays hole-like carriers (\ie a positive $R_H$) up to 200 K~\cite{Lee:2007},  9\% Fe substitution flips the sign of $R_H$, implying the majority carriers are electron-like below 20 K.  This is consistent with a simple band picture where $n_{\eff} = 1/(R_H |e|)$, and addition of electrons via Fe doping  changes the majority carrier type.  Similar behavior is observed in the 13\% sample.  Both recover the positive linear $B$ dependence of $\rho_{yx}$ at  $T \gg T_C$, implying the dominance of positive extrinsic carriers at high $T$. 

On the other hand, the Hall signal in 6\% exhibits a complex dependence on $T$ and $B$, which makes its division into the normal, anomalous, and topological Hall effects challenging.  Detailed $\rho_{yx}$ traces for the 6\% sample are shown in the supplementary information.

We observe that in addition to an increase of the topological Hall effect, substitution of Fe also increases the anomalous Hall effect (AHE), consistent with the concurrent increase of the longitudinal resistivity $\rho$.  The sign of the AHE changes with Fe content as well: it is negative for 0\%~\cite{Lee:2007} and 6\%, and positive for 9\% and 13\% (see supplementary information).  When $T \ll T_C$,  $\rho_{yx}$ varies in direct accordance with $M(B)$ (see Fig.~\ref{rhoM}(b)), which implies the AHE dominates in this regime.  The clear kink at low $T$s marks the spin alignment field $B_p \approx 0.6$ T at which all spins become collinear, forcing $b_r$ to be 0.  At intermediate $T$, the normal, anomalous, and topological Hall effects all contribute to $\rho_{yx}$, and it is in general non-trivial to distinguish them, which we discuss shortly.

The striking differences in $\rho_{yx}$ between $P$ and doping are 
(i) The AHE is significantly smaller under $P$, consistent with the reduction in $\rho$.
(ii) The THE is larger under $P$, relative to doped samples with comparable $T_C$s.  This is sensible, given that $m_s$ values are higher under pressure than in comparable doped samples, and a larger $m_s$ increases the overall Hund interaction between the spin texture and conduction electrons.
(iii) For MnSi under $P$, the region in the $T$-$B$ plane where the THE appears extends to lower $T$s than  in comparable doped samples.  Still, a similar broadening in the $B$ dimension occurs both in the 13\% Fe doped sample and in pure MnSi at 13 kbar (see Fig.~\ref{Hall}(c),(f) and  Ref.~\cite{Ritz:2013}). Such an extension has been observed in 2D FeGe films~\cite{Yu:2010}, where low dimensionality enhances fluctuations, and helps stabilize the spin texture in a larger range of $T$ and $B$.  In our system, suppression of $T_C$ may play a similar role. 

To parse the different Hall effects in Mn$_{1-x}$Fe$_x$Si, we combined measurements of the magnetoresistance (MR) and $M$ over field sweeps of $\pm$5 T.  For 9\% and 13\%, this allowed fittings of  $\rho_{yx}$ with the expression
\be
\rho_{yx}(B,T) = R_HB+ S_H\rho^2 M\Big(1+\alpha\frac{\rho_0}{\rho}\Big) +\rho_{yx}^T,
\label{AHEeq}
\ee
as in Ref.~\cite{Lee:2007}.  Here $R_H$  and $S_H$ are the normal and anomalous Hall coefficients, and the fitting parameters are $R_H$, $S_H$, and $\alpha$.  $\rho_0$ denotes $\rho(B=0)$.  The first two terms correspond to $\rho_{yx}^N$ and  $\rho_{yx}^A$.  Note that the second term $\rho_{yx}^A$ contains both intrinsic contributions to the AHE, where the anomalous Hall conductivity $\sigma_{xy}^A$ is linearly proportional to $M$ \cite{Lee:2007}, as well as extrinsic contributions to the AHE, where $\rho_{yx} \propto \rho$ \cite{Nagaosa:2010,Checkelsky:2008}.  The fitting parameter $\alpha$ characterizes this proportionality, which was found to be zero for MnSi \cite{Lee:2007}.

\begin{figure}[htb]
\begin{center}
  \includegraphics[width=0.95\linewidth]{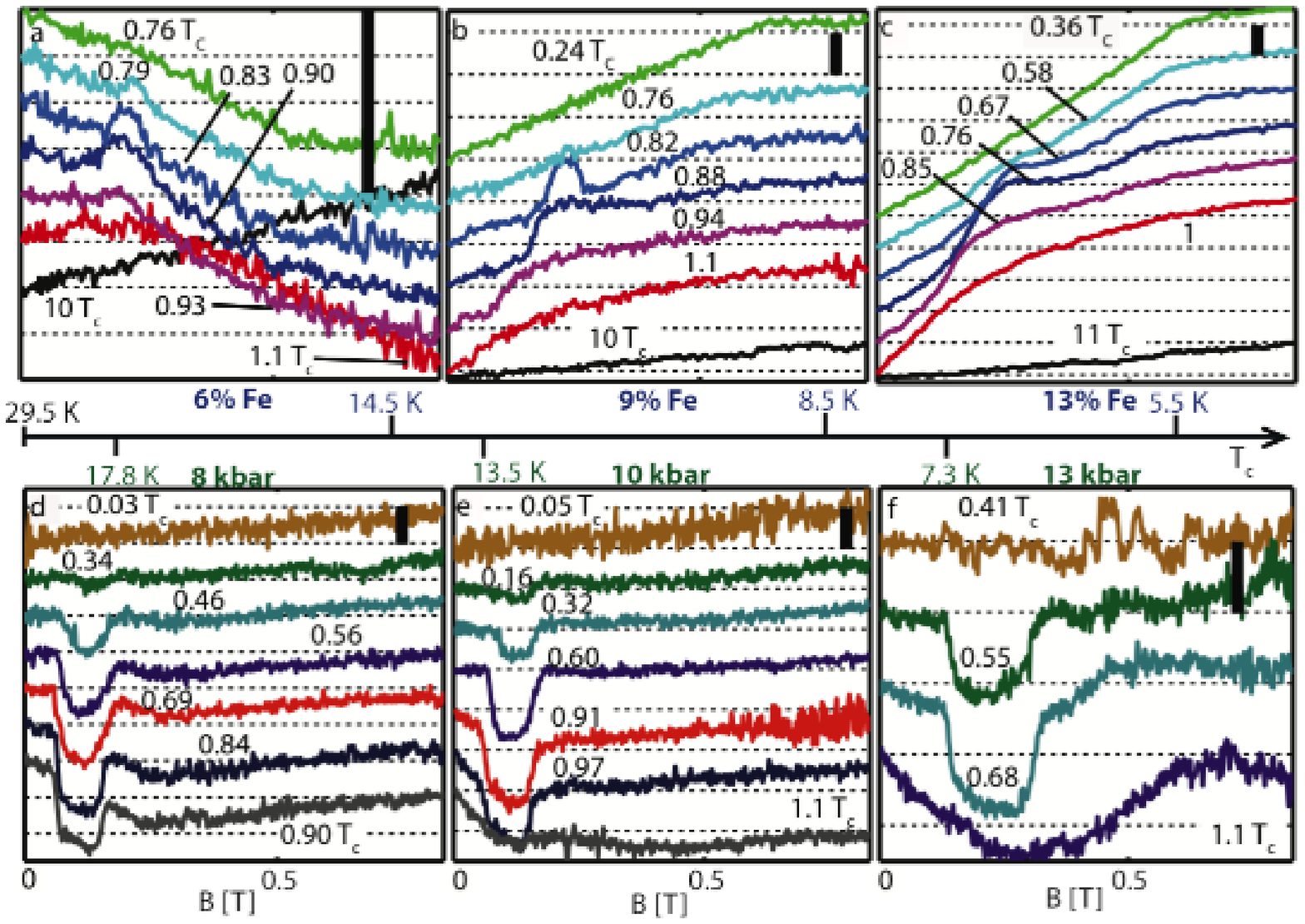}
  \caption {\small  $\rho_{yx}$ as a function of $B$ for Mn$_{1-x}$Fe$_x$Si (a)-(c) and MnSi under pressure (d)-(f). Black bars indicate scale of 20 n$\Omega$cm.  Different traces correspond to different fixed $T$s in units of $T_C$.  The temperature scale in the center shows how pressure and doping suppress $T_C$ for each dataset.  Curves are offset for readability.
   %: 0.76, 0.79, 0.83, 0.9, 0.93, 1.1, and 10 $T_C$ (6\%); 0.24, 0.76, 0.82, 0.88, 0.94, 1.1, and 10 $T_C$ (9\%); 0.36, 0.58, 0.67, 0.76, 0.85, 1, and 11 $T_C$ (13\%); 0.03, 0.34, 0.46, 0.56, 0.69, 0.84, and 0.90 $T_C$ (8 kbar); 0.05, 0.16, 0.32, 0.60, 0.91, 0.97, and 1.12 $T_C$ (10 kbar); 0.41, 0.55, 0.68, and 1.10 $T_C$ (13 kbar).%0.24, 0.47, 0.71, 0.95, and 1.19 $T_C$ (13 kbar).--this commented list is for pressure falling data
}
\label{Hall}
\end{center}
\efig

The topological contribution $\rho_{yx}^T$ is obtained by subtracting $\rho_{yx}^N$ and $\rho_{yx}^A$ (estimated from fitting) from the measured $\rho_{yx}$ shown in Fig.~\ref{Hall}.  For this reason our fitting results are focused on the $A$-phase.  To estimate the topological signal for 6\%, where the Hall signal is not described by Eq.~\ref{AHEeq} (see supplementary information), a linear background was subtracted from the Hall traces, as in \cite{Neubauer:2009,Ritz:2013}.  The pressure data were analyzed in the same way.

Table~\ref{t2} summarizes our results on the THE, calculations of the emergent magnetic field $B_{\eff}$, and comparison with theoretical upper bounds, for both Fe doping and $P$. The areas used to calculate $\bar{b_r}$ were extracted from interpolations of the helical wave-vector measured with neutron scattering on samples of nearby Fe content \cite{Grigoriev:2009}. $B_{\eff}$, calculated from the measured $\rho_{yx}^{T}$ and $R_H$, is expected to be always smaller than $\bar{b_r}$ (a theoretical upper-bound) such that $B_{\eff} = f \bar{b_r}$ with $0 < f < 1$, and this is observed.  Interestingly, despite the larger size of the THE in MnSi under $P$, $f$ remains more or less constant across both doping and pressure.  We suggest three physical considerations that will be reflected in the value of $f$: 
(i) The strength of the coupling between charge carriers and the spin texture.  Only in the strong Hund coupling limit, where the spin of conduction electrons tightly follows the spin-texture, can $f$ approach unity. 
(ii) The ratio of conduction electrons with majority to minority spins.  This can be varied by changes in the band structure through addition of Fe. Such alterations change the magnitude of the THE, as electrons with opposite spin feel opposite emergent fields.  These induce opposite Hall voltages which cancel with one another.
(iii) The presence of strong fluctuations in the spin texture on much shorter~\cite{Uemura:2006} time scales than our measurement (ms).  This is expected to reduce the time-averaged value of the emergent field $B_{\eff}$, and hence diminish $f$.

Another observation from our study on Fe doped MnSi is the identification of a positive MR which is prominent against a background of negative MR only for temperatures and magnetic fields where the THE appears.  This clearly shows that longitudinal electric transport is also sensitive to the presence or absence of SLs in Mn$_{1-x}$Fe$_x$Si.

Fig.~\ref{MR} (a)-(c) shows the magnetoresistance at various fixed $T$s.  Traces at 2K indicate a clear kink at $B_p$, attributed to a reduction of spin scattering in the transition from a conical to spin-collinear configuration.  As $T$ increases,  thermal fluctuations decrease the efficacy of the Zeeman coupling, and $B_p$ decreases slightly.

\begin{figure}%[ht]
\begin{center}
\includegraphics[width=1 \linewidth]{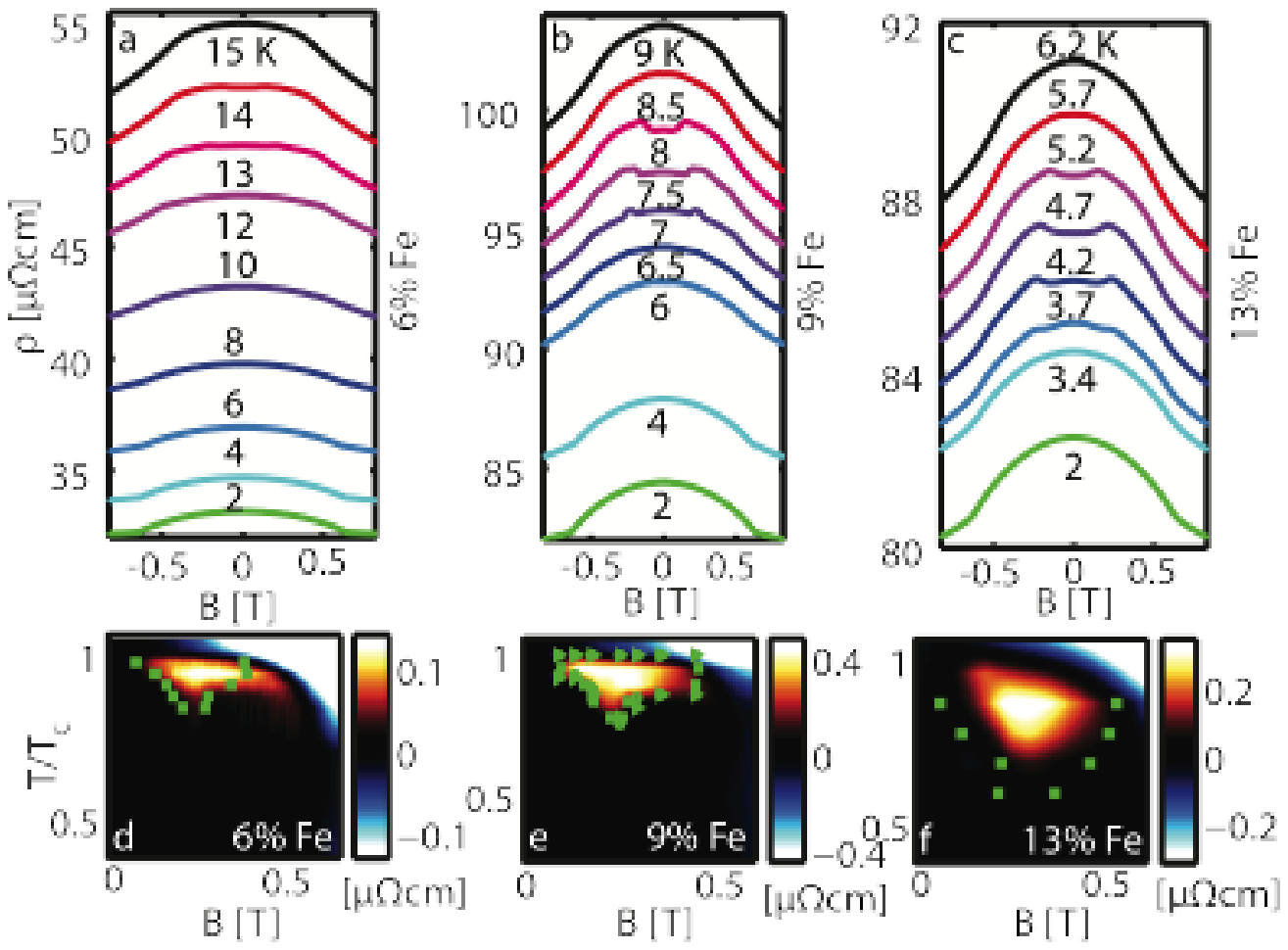}
\caption{\small (a)-(c): Magnetoresistance $\rho_{xx}=\rho$ at fixed temperatures.  (d)-(f): MR after subtracting the background from the conical phase (see text).  The difference is non-zero only in the $A$-phase, the boundaries of which are confirmed by the presence of the THE in sweeps of $T$ (triangles and circles) and $B$ (squares).}
 \label{MR}
\end{center}
\end{figure}

As $T$ approaches $T_C$, however, the $B$ profile of the MR changes to yield a \emph{positive} MR within a narrow range of $B$,  which coincides well with the range of $B$ where the THE appears.  The feature is directly visible in the 9\% and 13\% traces in Fig.~\ref{MR}: as $T$ approaches $T_C$ from below, a horn-like feature begins to emerge, then becomes wider as $T$ increases.  It vanishes and the MR recovers a smooth and monotonically decreasing dependence on $B$ when $T>T_C$.  We note that a similar horn-like feature, which disappears above $T_C$, was also observed in MnSi under $P$ \cite{Lee:2009,Ritz:2013}.

To make the feature more discernible, we empirically estimate the background of  the spin scattering reduction to be proportional to $B^2$, thus the horn-like feature $\Delta\rho^T(B)$ is estimated as $\Delta\rho^T(B)  = \rho(B) - \left[\rho_0 - aB^2\right]$, where $\rho_0= \rho(B=0)$ and $a$ is a positive and weakly $T$-dependent fitting parameter. 

In Fig.~\ref{MR} (d)-(f), we plot $\Delta\rho^T$ in the $T$-$B$ plane with the magnitude of $\Delta\rho^T$ indicated by the color scale.  The plots delineate a region in the $T$-$B$ plane which agrees with the boundaries set by the THE, in sweeps of both $T$ (triangles, warming; circles, cooling) and $B$ (squares).

In summary, measurements of the THE in Mn$_{1-x}$Fe$_x$Si have revealed the tens-of-Tesla emergent magnetic fields generated by its exotic spin texture, and showed they can be enhanced in a controlled manner with Fe doping.  As $T_C$ is suppressed with increasing $x$, we observe changes in the sign and magnitude of the THE, consistent with an enhancement of the gauge field by the reduction of the magnetic unit cell and the sign change of $R_H$.  
Moreover, a positive MR in Fe doped samples was observed only within a narrow region of the $T$-$B$ plane, which coincides well with the region where the THE appears.  While the magnitude of THE is substantially larger under $P$ than in doped samples, the ratio $f$ of the effective field to its theoretical maximum remains comparable in both cases. 

 %{\color{red} This leaves major open questions about the factors that determine $f$; chief among these is the effect of pressure and iron doping on the spin polarization of charge carriers.}  Further studies are needed to improve this understanding and allow fine manipulation of both $f$ and the enormous gauge fields themselves.

\vspace{0.1in}
\noindent{\emph {Acknowledgement}} The authors thank  M. Hermele, D. Reznik, and J. A. Yang for enlightening discussions. This work is supported by the U.S. Department of Energy, Office of Basic Energy Sciences, Division of Materials Sciences and Engineering under Award ER 46797.

\vspace{0.1in}
\noindent$^\dagger$ Current Address: Department of Physics and Astronomy, University of California, Riverside CA 

%\bibliography{MnSiBib}

\end{document}